\documentstyle[epsfig,aps,preprint]{revtex}
\newcommand{\n}{\noindent}
\tighten
\begin{document}

\title{Recurrence of fidelity in near integrable systems}

\author{R. Sankaranarayanan \footnote {Present Address: Department of Physics
of Complex Systems, Weizmann Institute of Science, 76100 Rehovot, Israel.}
and Arul Lakshminarayan \footnote {Present Address: Department of Physics,
Indian Institute of Technology, Madras, Chennai 600036, India.}}
\address{Physical Research Laboratory, Navrangpura, Ahmedabad 380009, India.}
\maketitle

\begin {abstract}
Within the framework of simple perturbation theory, recurrence time of
quantum fidelity is related to the period of the classical motion. This
indicates the possibility of recurrence in near integrable systems. We have
studied such possibility in detail with the kicked rotor as an example. In
accordance with the correspondence principle, recurrence is observed when
the underlying classical dynamics is well approximated by the harmonic
oscillator. Quantum revivals of fidelity is noted in the interior of
resonances, while classical-quantum correspondence of fidelity is seen to
be very short for states initially in the rotational KAM region.

\end {abstract}

\section {introduction}

Qualitative dynamical behaviour of a classical system is characterized by the
sensitivity of its orbits with respect to initial conditions. The sensitivity
also changes with parameters of the system. If an orbit is not sensitive to
initial conditions, the underlying dynamics is regular. On the other hand, for
chaotic dynamics the orbit is highly sensitive to initial condition. Lyapunov
exponents, which measure the rate of exponential divergence of two initially
close orbits, is zero for the regular case and positive for the chaotic case.
An analogous classification of dynamics is not possible in the quantum domain,
since quantum theory does not accommodate the very concept of orbits. 

If we associate two localized wave packets which are initially close-by in
state space, overlap of these states remain invariant under unitary quantum
evolution. The corresponding classical Liouville evolution of phase space
densities is also a linear unitary one, but the ability to develop structures
on infinitesimal scales, without the restrictions of the uncertainty
principle, coexists with underlying orbit chaos. In the quantum mechanical
case exponential instability of for instance the wave packet center is seen
for quantized chaotic systems over a short time scale, corresponding to the
Ehrenfest time scale. 

One approach to quantify quantum sensitivity was initiated a while back by 
Peres \cite{peres84} who proposed the overlap of states that are evolved
from the same initial state $|\alpha\rangle$ under two Hamiltonians differing
by a small perturbation. The overlap intensity, also known as fidelity, is
defined as

\begin{equation}
f(t) = |\langle\alpha|e^{iH_1t/\hbar}e^{-iH_0t/\hbar}|\alpha\rangle|^2
\end{equation}

\n where $H_0$ represents the Hamiltonian of an unperturbed system, 
$H_1=H_0+\epsilon V$ represents the perturbed system and $\epsilon$ is
the perturbation parameter. Fidelity has evoked enormous interest in recent
years with different interpretations. It is a measure of quantum 
irreversibility in the context of decoherence when the system under
investigation interacts with a classically chaotic environment \cite{jala01}.
It is realized as polarization echo in nuclear magnetic resonance,
measuring the irreversibility in many body quantum dynamics \cite{expt}.
It also characterizes the loss of phase coherence in quantum
computation \cite{nielsen}. 

Initial investigations by Peres \cite{peres84} show that $f(t)$ is appreciable
(on a time average) and its fluctuations are more for the regular case than
for the chaotic case. Some of the recent studies have focused on the decay
rate of $f(t)$ for chaotic systems \cite{ph01}. More investigations on this
topic show that while the fidelity decays on time scale $\sim 1/\epsilon^2$
for chaotic systems, it decays on shorter time scale $\sim 1/\epsilon$ for
integrable systems \cite{prosen02}. In \cite{ph03} the decay for integrable
system is shown to have power-law behaviour.

In this paper, using perturbative approximations, we relate recurrence
time of the quantum fidelity to the classical period. This indicates that
recurrence of quantum fidelity would be possible in nearly integrable systems.
With the kicked rotor as a dynamical model, we examine this recurrence
phenomena in detail. Our numerical experiments also reveal that the evolution
of fidelity exhibits a wide range of behaviour, depends on the choice of
initial state. 

\section {formulation}

Let us consider an unperturbed system $H_0 \equiv H(\omega)$, with $\omega$
being the system parameter, satisfying the eigenvalue equation

\begin{equation}
H_0|u_n\rangle = E_n|u_n\rangle \, .
\end{equation}

\n Let $H_1 \equiv H(\omega^\prime)$ be the perturbed system where
$\omega^\prime = \omega + \epsilon$ and $\epsilon$ is the perturbation
parameter. Under a small perturbation, for the simplest zeroth order
approximation to the fidelity it is sufficient to assume that the energy
eigenvalues differ up to the first order of $\epsilon$ and the corresponding
eigenstates remain unaffected.  It is easy to see that the eigenstate changes
contribute to the fidelity at an order $\epsilon$.

That is the approximation we can make for the calculation of fidelity 
amounts to the approximation:
 
\begin{equation} 
H_1|u_n\rangle \approx \tilde E_n|u_n\rangle \;\; ; \;\;
\tilde E_n \approx E_n + \epsilon\;{dE_n\over d\omega} \, . 
\end{equation}

\n Here $dE_n/d\omega$ measures the rate at which $n$th level changes
with the parameter. It is also known in the literature as level
velocity wherein the parameter is thought of as pseudo-time. In this
approximation the fidelity can be written as

\begin{equation}
f(t) \approx \left|\sum_n |c_n|^2 \exp\left[-{i\epsilon t\over\hbar}
\left({dE_n\over d\omega} \right)\right]\right|^2
\end{equation} 

\n where $c_n = \langle u_n|\alpha\rangle$. Note that $f(0)=1$ and we are
interested in the time evolution of the fidelity. Assuming that the weighting
probabilities $|c_n|^2$ are strongly centered around a mean $\bar n$, one can
expand $dE_n/d\omega$ using Taylor series in $n$ around $\bar n$ up to the
first-order correction as

\begin{equation}
{dE_n\over d\omega} = {dE_{\bar n}\over d\omega} + (n-{\bar n}) 
{d^2E_{\bar n}\over dn\, d\omega} \; . 
\end{equation}

\n This gives the recurrence time 

\begin{equation}
t_r = \left({2\pi\hbar \over\epsilon}\right)
\left|{d^2E_{\bar n}\over dn\, d\omega}\right|^{-1}
\end{equation}

\n such that the fidelity becomes unity when $t$ is an integer multiples 
of $t_r$.
Recollecting that the classical action $S$ is quantized as $(2\pi\hbar)n = S$,

\begin{equation}
t_r = {1\over\epsilon}
\left|{d^2E_{\bar n}\over dS\, d\omega}\right|^{-1} \, .
\end{equation}

\n In terms of $T = dS/dE$, the period of the classical orbit
underlying the coherent state $|\alpha\rangle$, the recurrence time is
thus given by a combination of simple perturbative and semiclassical
approximations as

\begin{equation}
t_r = {T^2\over\epsilon}\left|{dT\over d\omega}\right|^{-1} .
\end{equation} 

\n Thus the recurrence time of quantum fidelity is directly related to
period of the underlying classical motion as well as to its
variability to the relevant parameters.

\section {harmonic oscillator}

According to the above relation, the quantum fidelity for the 
harmonic oscillator 

\begin{equation}
H(\omega) = {p^2\over 2} + {1\over 2} {\omega}^2q^2
\end{equation}

\n with a classical period $T=2\pi/\omega$ recurs at

\begin{equation}
t_r = {2\pi\over\epsilon} \, .
\label{tr}
\end{equation}

\n This recurrence time can also be obtained from the underlying classical
dynamics by evolving an initial phase space density $\rho\,(q,p)$ which
satisfies the normalization condition

\begin{equation}
\int\rho\,(q,p)\,dq\,dp = 1 \, .
\end{equation}

Analogous to the quantum fidelity, we define normalized classical fidelity as

\begin{equation}
F(t)= \left\{\int\rho^2\,(q,p)\,dq\,dp\right\}^{-1} 
\int \rho\,(q,p)\,\rho\,(q_{2t},p_{2t}) \,dq\,dp \, . 
\label{fc}
\end{equation} 

\n where $\rho\,(q_{2t},p_{2t})$ is the final phase space density which is
obtained from forward time evolution of the initial density for a time $t$
with the frequency being $\omega$, followed by backward time evolution for
the same time $t$ with the frequency set as $\omega^\prime$. The classical
fidelity measures the overlap of the initial density with the final density
so obtained.

If the forward time evolution of the oscillator is given by $(q,p) \rightarrow
(\tilde q,\tilde p)$ i.e., 

\begin{equation}
\left[ \begin{array}{c} \tilde q \\ \tilde p \end{array} \right] =
\left[ \begin{array}{cc} \cos(\omega t) & (1/\omega)\sin(\omega t) \\ 
-\omega\sin(\omega t) & \cos(\omega t) \end{array}\right]  
\left[\begin{array}{c} q \\ p \end{array}\right] \, ,
\end{equation}

\n the initial density evolves forward in time $t$ as $\rho\,(q,p) \rightarrow
\rho\,(q_t,p_t)$ where $q_t$ and $p_t$ are given by the relation:

\begin{equation}
\left[\begin{array}{c} q_t \\ p_t \end{array}\right] =
\left[\begin{array}{cc} \cos(\omega t) & -(1/\omega)\sin(\omega t) \\ 
\omega\sin(\omega t) & \cos(\omega t) \end{array}\right]  
\left[\begin{array}{c} q \\ p \end{array}\right] \, . 
\label{forward}
\end{equation}

\n Then we consider the backward evolution for the same time $t$ 
with the frequency now set to $\omega^\prime$;
$\rho\,(q_t,p_t) \rightarrow \rho\,(q_{2t},p_{2t})$ where $q_{2t}$ and
$p_{2t}$ are given by the relation: 
 
\begin{equation}
\left[\begin{array}{c} q_{2t} \\ p_{2t} \end{array}\right] =
\left[\begin{array}{cc} \cos(\omega^\prime t) & 
(1/\omega^\prime)\sin(\omega^\prime t) \\ 
-\omega^\prime\sin(\omega^\prime t) & \cos(\omega^\prime t) \end{array}\right]  
\left[\begin{array}{c} q_t \\ p_t \end{array}\right] \, . 
\label{backward}
\end{equation}

\n We shall now write  

\begin{equation}
\left[\begin{array}{c} q_{2t} \\ p_{2t} \end{array}\right] =
\left[\begin{array}{cc} a & b \\ c & d \end{array}\right]  
\left[\begin{array}{c} q \\ p \end{array}\right] 
\end{equation}

\n where the elements of square matrix are obtained by combining the Eqns.
(\ref{forward}) and (\ref{backward}). 

If $q_{2t}=q$ and $p_{2t}=p$ for some time $t=t_R$, then $F(t_R) = 1$ and
the classical fidelity recurs fully. This condition is given by

\begin{equation}
\left|\begin{array}{cc} a-1 & b \\ c & d-1 \end{array}\right| = 0 \, .
\end{equation}

\n Since $(ad-bc)=1$, the above condition becomes $a+d=2$. The recurrence
time $t_R$ then satisfies the relation 

\begin{equation}
2\cos(\omega t_R)\cos(\omega^\prime t_R) + 
\left[{\omega^2+\omega^{\prime 2}\over \omega\omega^\prime}\right]
\sin(\omega t_R)\sin(\omega^\prime t_R) = 2 \, .
\end{equation}

\n Substituting $\omega^\prime = \omega+\epsilon$ and approximating
$\omega^{\prime 2}\approx \omega^2+2\epsilon\omega$ (up to the first order
of $\epsilon$), this condition reduces to 

\begin{equation}
\cos(\epsilon t_R)=1 \;\;\; \hbox{or} \;\;\; t_R = {2\pi\over\epsilon} \, .
\end{equation}

Thus recurrence time $t_R$ of the classical fidelity, at this level of
approximations, is identical to that of the quantum fidelity given by the
Eqn.(\ref{tr}). This shows that recurrence of quantum fidelity for the
harmonic oscillator is a manifestation of the underlying classical dynamics.
If we replace $\omega$ by $i\omega$ and $\omega^\prime$ by $i\omega^\prime$,
the dynamics corresponds to the unstable motion of an inverted harmonic
oscillator. In this case the recurrence condition is

\begin{equation}
2\cosh(\omega t_R)\cosh(\omega^\prime t_R) -
\left[{\omega^2+\omega^{\prime 2}\over \omega\omega^\prime}\right]
\sinh(\omega t_R)\sinh(\omega^\prime t_R) = 2 \, .
\end{equation}

\n Up to the first order of $\epsilon$, this condition reduces to
$\cosh(\epsilon t_R)=1$ which has only the trivial solution $t_R=0$. Hence
there is no recurrence in the classical fidelity of the unstable oscillator.

\section {kicked rotor}

For further investigations on the recurrence phenomena in near integrable
systems, we consider the kicked rotor whose Hamiltonian is 

\begin{equation}
H = {p^2\over 2} + V(q) \sum_j \delta (j-t) \, .
\end{equation}

\n Here $V(q) = k\cos(2\pi q)/4\pi^2$ and $k$ is the kick strength - the only
parameter of the system. The corresponding kick-to-kick dynamics is given by
the standard map: 

\begin{equation}
\left. \begin{array}{lll}
p_{t+1}&=&p_t+(k/2\pi)\sin(2\pi q_t ) \\[6pt]
q_{t+1}&=&q_{t}+p_{t+1}
\end{array}\right\} (\hbox{mod}\;1) \, .
\end{equation}

\n Note that the modulo operation restricts $q$ and $p$ to between $-1/2$ and
$1/2$ with the opposite edges being identified. The kick-to-kick
quantum dynamics is given by the quantum map: 

\begin{equation}
|\psi_{t+1}\rangle = U_0|\psi_t\rangle \;\; ; \;\;
U_0 = \exp[-ip^2/2\hbar]\,\exp[-iV(q)/\hbar]
\end{equation}

\n where $U_0 \equiv U(k)$ is the unit-time quantum propagator. 

The 2-torus phase space can be quantized upon introducing periodic boundary
conditions in both the canonical variables \cite{ford91}. This imposes finite
number, $N$, of quantum states such that $N=1/2\pi\hbar$ ($N\rightarrow\infty$
is the classical limit); we take $N=500$ for the following calculations. Since
the Hamiltonian is time periodic, according to Floquet theory \cite{jordan}
fundamental solutions statisfy the eigenvalue equation

\begin{equation} 
U_0|\phi_n\rangle = e^{i\phi_n}|\phi_n\rangle
\end{equation}

\n where $\phi_n$ are real (between 0 and $2\pi$). If $\phi_n = E_n/\hbar$,
then $E_n$ are called quasienergies which are analogous to the energy eigenvalues
of the time-independent Hamiltonian system \cite{sambe73}. Correspondingly the
states $|\phi_n\rangle$ are called quasienergy states. It is worth noting that
as a consequence of hermiticity of the Hamiltonian, the quasienergy states
$|\phi_n\rangle$ are orthogonal and they form a complete set in the $N$
dimensional space. The quasienergy spectrum can be numerically obtained by
diagonalizing the matrix representation of $U$.  Choosing the discrete position
eigenstates $|l\rangle$ as the basis, $U$-matrix takes the form \cite{arul97} 

\begin{equation}
\langle l|U|l^\prime\rangle = {1\over \sqrt{N}}
\exp \left[-i\pi\left\{{1\over 4} - {{(l-l^\prime)}^2\over N} + 
2NV\left({l^\prime\over N}\right) \right\} \right] 
\end{equation}

\n where $l,l^\prime = -N/2,-N/2+1,\ldots N/2-1$. 

Let us consider $U_0$ as an unperturbed system and $U_1 \equiv U(k+\delta k)$
as the perturbed system. Then the quantum fidelity is given by

\begin{equation}
f(t) = |\langle\alpha|U_1^{-t}U_0^t|\alpha\rangle|^2 .
\end{equation}

\n In accordance with our earlier approximations the eigenstates of
the perturbed system are taken to be such that

\begin{equation} 
U_1|\phi_n\rangle \approx e^{i\tilde\phi_n}|\phi_n\rangle \;\; ; \;\;
\tilde\phi_n \approx \phi_n + \delta k {d\phi_n\over dk}
\end{equation}

\n where $d\phi_n/dk = \hbar^{-1}\langle\phi_n|dV/dk|\phi_n\rangle$, quantum
fidelity for the kicked rotor can be approximated as

\begin{equation}
f(t) \approx \left|\sum_n |\langle\phi_n|\alpha\rangle|^2 
\exp\left[-{it\delta k\over\hbar}\left(d\phi_n\over dk
\right)\right]\right|^2 \, . 
\label{fp}
\end{equation}

\n This gives the recurrence time as

\begin{equation}
t_r = {T^2\over\delta k}\left|{dT\over dk}\right|^{-1} .
\end{equation} 

\n with $T$ being the classical period of the rotor.

\section {numerical results}

\begin{figure}
\centerline{\psfig{figure=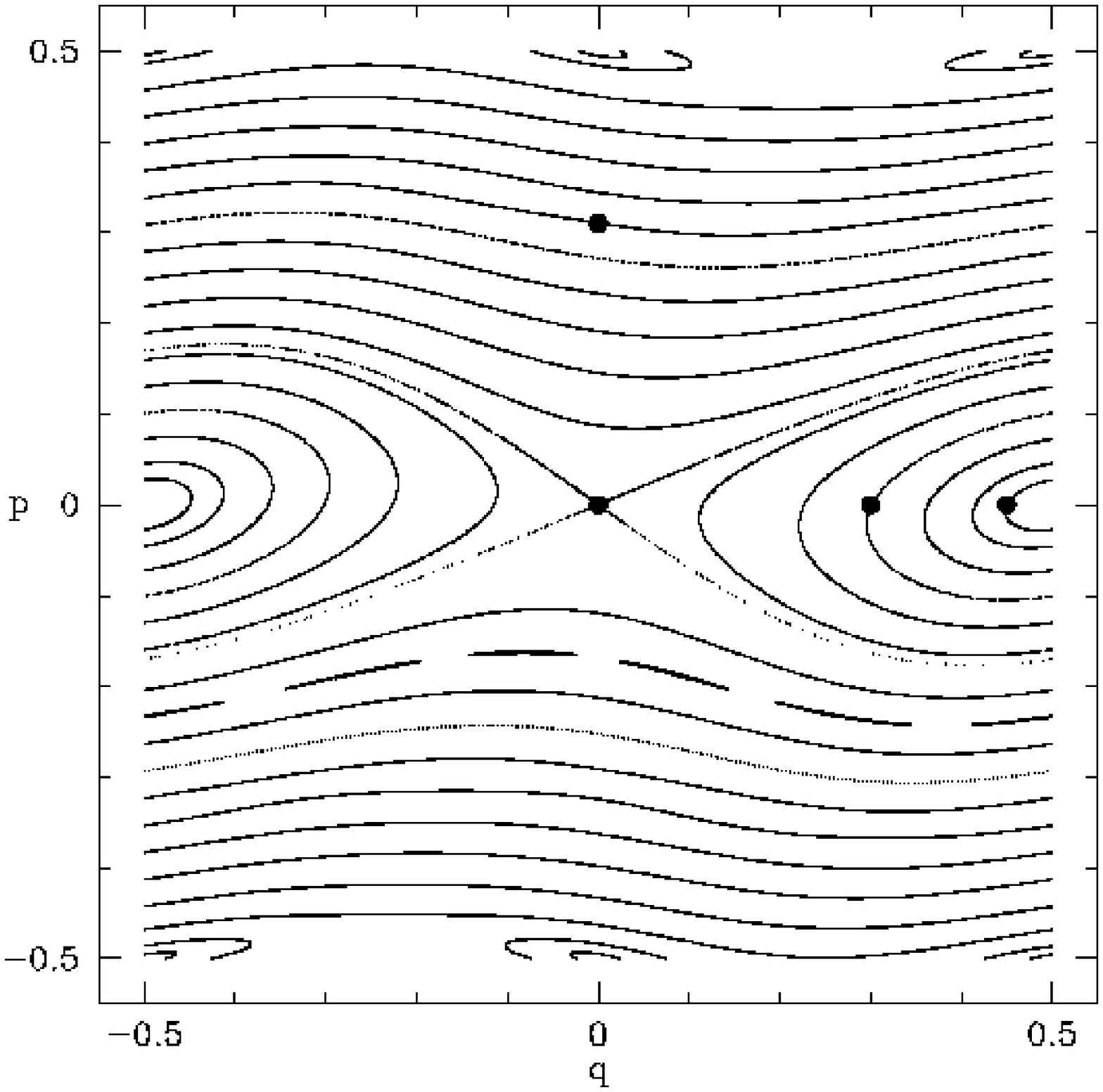,height=10cm,width=10cm}}
\caption{Phase space portrait of the standard map for $k=0.3$. The points
$(q_0,p_0)=(0.45,0),(0.3,0),(0,0)$ and $(0,0.3)$, shown as solid circles,
correspond to the intitial coherent states used for the quantum evolution.}
\label{xp}
\end{figure}

The dynamical features of the standard map depends on the kick strength.
In the absence of kick ($k=0$), it is just the twist map with regular
dynamics. For small kick strengths ($k<1$), the phase space is predominantly
regular with primary nonlinear resonance zones along $p=0$ and there are
large number of smooth Kolmogorov-Arnold-Moser (KAM) tori in large $|p|$
regions on which the motion is quasiperiodic (see Fig. \ref{xp}). These two
distinct phase space regions are demarcated by a separatrix. In this case,
the phase space is similar to that of the simple pendulum. As $k$ increases,
the KAM tori are slowly destroyed and at $k\approx 1$ all the KAM tori are
destroyed \cite{greene79} leading to the onset of global chaos. When $k\gg 1$,
the dynamics is predominantly chaotic. In this paper, we take $k=0.3$
and $\delta k = 0.01$ throughout, as our interest is restricted to the near
integrable regime.

To evaluate the quantum fidelity for different regions of the phase space, we
take the initial state as a coherent state peaked at $(q_0,p_0)$, i.e.,
$|\alpha\rangle = |q_0,p_0\rangle$. It is a minimum uncertainty (very nearly
Gaussian) wave packet with equal spread $\sigma = \sqrt{\hbar/2}$ in both the
canonical variables. Such an initial state on the torus can be constructed
by the method devised by Saraceno \cite{sar90}. The initial states we have
chosen for the following computations are marked in Fig. \ref{xp}.

For comparison, we also compute the classical fidelity by taking an initial
phase space density to be a Gaussian 

\begin{equation}
\rho\,(q,p) = {1\over 2\pi\sigma^2} \exp\left[-{1\over 2\sigma^2}
\left\{(q-q_0)^2 + (p-p_0)^2\right\}\right] \, .
\end{equation}

\n This density is a classical equivalent of the coherence state
$|q_0,p_0\rangle$ containing an ensemble of $10^4$ initial conditions. 
For the classical evolution, each point of the initial density is forward
iterated using the standard map (with parameter $k$) for $t$ steps, followed
by backward iteration for $t$ steps by the corresponding time reversed map,
with $k$ being replaced by $k+\delta k$. This gives the final density
$\rho\,(q_{2t},p_{2t})$. With this, we use a discretized version of
Eqn. (\ref{fc}) to calculate the classical fidelity.

It should be emphasized that the recurrence of fidelity is possible only
when the perturbative approximation is valid. In order to validate this
approximation for different regions of the phase space, we compute phase
space representation of inverse participation ratio (IPR) as 

\begin{equation} 
z(q,p) = \sum_n|\langle\phi_n|q,p\rangle|^4
\end{equation} 

\begin{figure}
\centerline{\psfig{figure=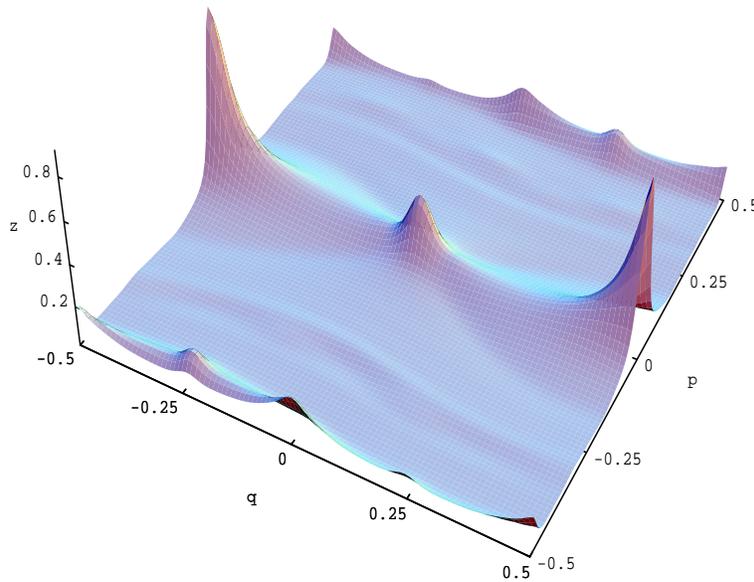,height=8cm,width=10cm}}
\caption{The phase space representation of the Inverse Participation
Ratio $z(q,p)$ plotted for $k=0.3$.}
\label{ipr}
\end{figure}

\n such that $z^{-1}$ measures the effective number of states $|\phi_n\rangle$
required to construct a given region of phase space.  As we see from
Fig. \ref{ipr}, while very few states are required to construct the region
near the stable fixed point, many states contribute to construct the
separatrix region and rotational KAM region. In other words, low energetic
part of the spectrum that corresponds to the neighborhood of stable fixed point
has very low density of states; high energetic part of the spectrum that 
corresponds to the separatrix and KAM regions are highly dense. Generally,
such simple perturbative expansions work badly for the dense spectrum. Hence
the approximation is expected to be valid near the stable fixed point, and
so is the recurrence of quantum fidelity. For other cases, the approximation
would not be sufficient and the recurrence is not expected. As we demonstrate
below, this is indeed the case. Nevertheless, for the dense spectrum,
``quantum revivals'' of fidelity may occur which is beyond the scope of
present study, but it is under active study.

To begin with, we focus on the neighborhood of the stable fixed point. Note
that Jacobian of the standard map evaluated at $(0.5,0)$ has eigenvalues
$e^{\pm i\lambda}$ and $\lambda$ is obtained from the relation

\begin{equation}
e^{i\lambda} + e^{-i\lambda} = 2\cos\lambda= 2-k \, . 
\end{equation}

\n Since $\lambda$ is the frequency of the oscillatory motion in this region,
the corresponding period is 

\begin{equation}
T(k) = {2\pi\over\lambda} = 2\pi\left[\cos^{-1}\left({2-k\over 2}
\right)\right]^{-1} 
\end{equation}

\n and we obtain the recurrence time as 

\begin{equation}
t_r = {2\pi\over\delta k} \sqrt{k(4-k)} \, .
\end{equation}

\n For the parameters we have considered here ($k=0.3$,$\;\delta
k=0.01$), the quantum fidelity recurs at $t_r = 662$. If we
approximate the underlying oscillatory motions by harmonic oscillators with
frequencies

\begin{equation}
\omega = \left[\cos^{-1}\left({2-k\over 2}\right)\right]^{-1} \;\; ; \;\;
\omega^\prime = \omega + \epsilon = \left[\cos^{-1}
\left({2-k-\delta k\over 2}\right)\right]^{-1} \, ,
\end{equation}

\n then $t_r=2\pi/\epsilon = 667$. 

\begin{figure}
\centerline{\psfig{figure=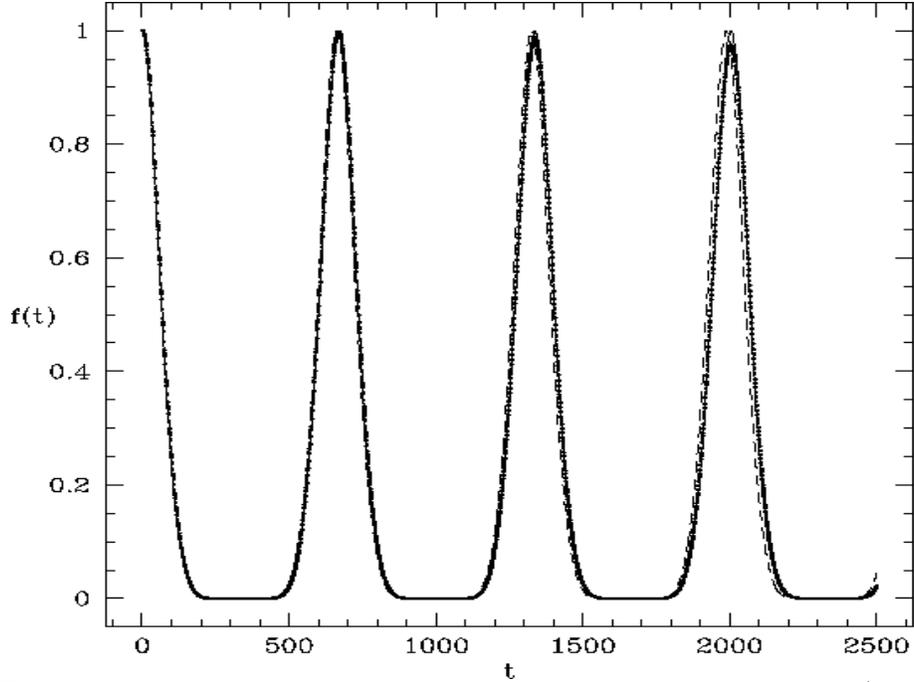,height=9cm,width=12cm}}
\caption{The solid line is the quantum fidelity for the initial state
centered at $(0.45,0)$, near the stable fixed point, and the dashed line
is the perturbative approximation (\ref{fp}). The corresponding classical
fidelity, evaluated from Eqn. (\ref{fc}), is shown as dots.}
\label{fid1}
\end{figure}

In Fig. \ref{fid1} we show the evolution of fidelity for the initial state
$|0.45,0\rangle$ that is localized in the neighborhood of the stable fixed
point. For initial time steps, quantum fidelity has a smooth fall-off from
unity to zero. It remains nearly zero for some time and then it rises and
recurs at $t=667$. This is close to the value 662 that is obtained from
perturbation theory. Notice that the recurrence time matches exactly with
that of the harmonic oscillator approximation. The fidelity also recurs
periodically and the agreement with perturbation theory is found to be good.
In addition, we find that the classical evolution exactly retraces quantum
evolution. Thus, fidelity for oscillatory motion in the neighborhood of a
stable fixed point exhibits very good quantum-classical correspondence.

\begin{figure}
\centerline{\psfig{figure=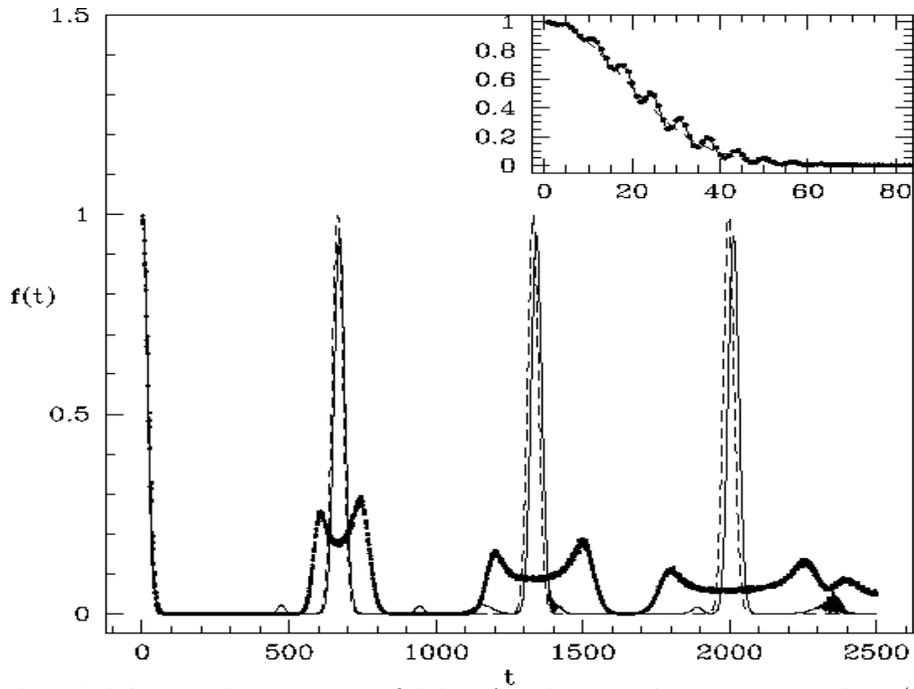,height=9cm,width=12cm}}
\caption{The solid line is the quantum fidelity for the initial state
centered at $(0.3,0)$, in the interior of the primary resonance zone,
and the dashed line is the perturbative approximation (\ref{fp}). For
comparison, the corresponding classical fidelity is shown as dots. Inset
shows the initial evolution.}
\label{fid2}
\end{figure}

Fig. \ref{fid2} corresponds to the state which is initially placed in the
interior of the primary resonance zone. In this case, the quantum fidelity
{\it almost} recurs i.e, $95\%$ recurrence is observed at $t=670$. Note that
this time is close to the recurrence time predicted from the harmonic
oscillator approximation. Here also the perturbative approximation agrees
fairly well with the actual evolution. However, unlike the previous case,
classical evolution initially follows the quantum evolution and then deviates
at later time steps. This shows that the observed {\it near} recurrence of
fidelity is a ``pure'' quantum phenomena. This shows interestingly that,
within the primary resonance zone, the quantum dynamics seems to be less
sensitive to nonlinearity.

\begin{figure}
\centerline{\psfig{figure=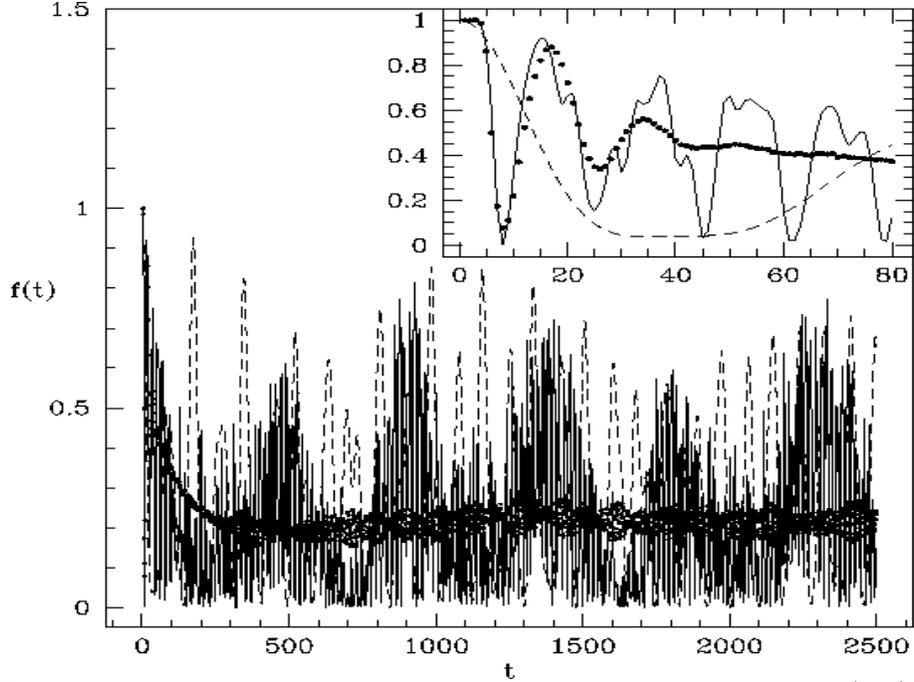,height=9cm,width=12cm}}
\caption{The solid line is the quantum fidelity for the initial state
centered at $(0,0)$, the unstable fixed point, and dashed line is the
perturbative approximation (\ref{fp}). Dots are the corresponding
classical fidelity and the initial evolution is shown in the inset.}
\label{fid3}
\end{figure}

As we see from Fig. \ref{fid3}, quantum evolution is highly complex when
the underlying classical motion is on the separatrix. The perturbative
approximation fails and the fidelity does not recur.  In this case,
asymptotically the classical fidelity becomes more stable in comparison to
the quantum counterpart. If we approximate unstable motion on separatrix by
an inverted harmonic oscillator, classical fidelity also does not exhibit
any recurrence.

\begin{figure}
\centerline{\psfig{figure=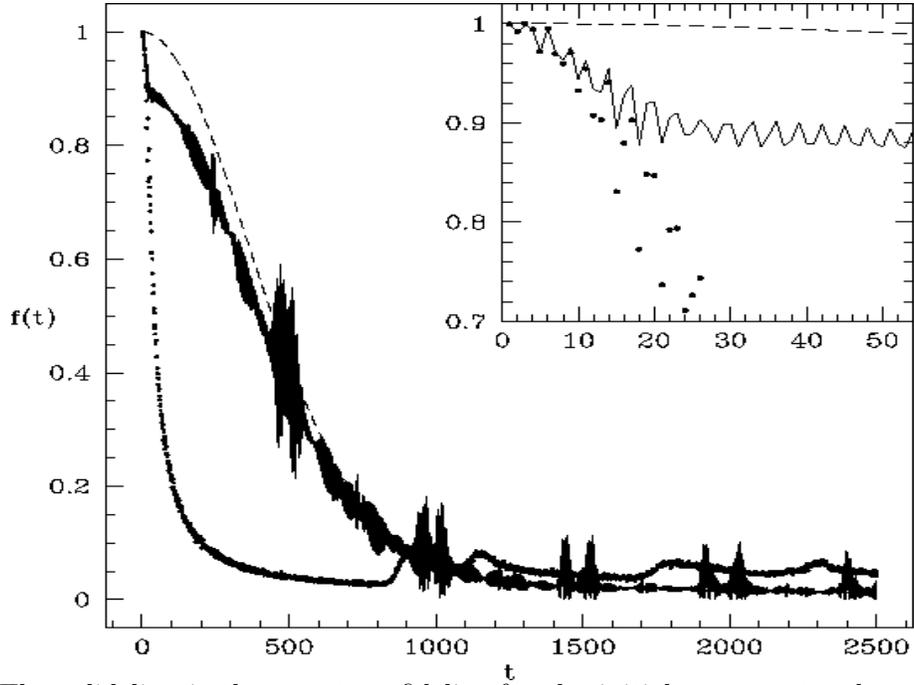,height=9cm,width=12cm}}
\caption{The solid line is the quantum fidelity for the initial state
centered at $(0,0.3)$, on a rotational KAM curve, and the dashed line is the
perturbative approximation (\ref{fp}). Dots are the corresponding
classical fidelity and the initial evolution is shown in the inset.}
\label{fid4}
\end{figure}

Finally, we move on to the rotational KAM region of the phase space. A
typical evolution of an initial state localized in this region is shown in
Fig. \ref{fid4}.  As expected, perturbative approximation fails and there are
no signatures of fidelity recurrence. Although classical motion on the KAM
tori is regular, we notice that quantum-classical correspondence is seen
only for a short time. This may be due to enhanced quantum interference in
this region of energetic rotational motion. More study on these issues is
currently underway.

\section {conclusion}

We have shown that, within the framework of simple perturbative and
semiclassical approximations, recurrence time of quantum fidelity is
related to period of the classical motion. In the case of the harmonic
oscillator, fidelity recurs at $t_r=2\pi/\epsilon$ where $\epsilon$ is
the perturbation parameter.  Invoking the kicked rotor as an example, we
have studied in detail the recurrence phenomena for nearly integrable
regimes. It is shown that quantum fidelity recurs only when the
underlying classical motion is well approximated by the harmonic
oscillator. This recurrence is also in accordance with the correspondence
principle. Though the recurrence is not possible in other cases, quantum
revival of fidelity can occur and this needs further investigation.


\hrule
\end{document}